\documentclass[3p,times,procedia]{elsarticle}
\usepackage{nupha_ecrc}

\volume{00}

\firstpage{1}

\journalname{Nuclear Physics A}

\runauth{}

\jid{nupha}

\jnltitlelogo{Nuclear Physics A}

\usepackage{amssymb}

\usepackage[figuresright]{rotating}

\begin{document}

\begin{frontmatter}



\dochead{XXVIth International Conference on Ultrarelativistic Nucleus-Nucleus Collisions\\ (Quark Matter 2017)}

\title{The QCD Equation of state and critical end-point estimates at $\mathcal O(\mu_B^6)$}


\author{Sayantan Sharma}

\address{Brookhaven National Laboratory, Upton, New York 11973\\
\textbf{[For Bielefeld-BNL-CCNU collaboration]}}

\begin{abstract}
We present results for the QCD Equation of State at non-zero chemical
potentials corresponding to the conserved charges in QCD using Taylor expansion 
upto sixth order in the baryon number, electric charge and strangeness
chemical potentials. The latter two are constrained by the strangeness neutrality
and a fixed electric charge to baryon number ratio. In our calculations, we
use the Highly Improved Staggered Quarks (HISQ) discretization scheme at
physical quark masses and at different values of the lattice spacings to control lattice cut-off effects. 
Furthermore we calculate the pressure along lines of
constant energy density, which serve as proxies for the freeze-out conditions
and discuss their dependence on $\mu_B$ , which is necessary for hydrodynamic
modelling near freezeout.
We also provide an estimate of the radius of convergence of the Taylor
series from the 6th order coefficients which provides a new constraint on
the location of the critical end-point in the T-$\mu_B$ plane of the QCD phase
diagram.
\end{abstract}


\end{frontmatter}


\section{Introduction}
\label{sec:intro}
There has been a considerable progress in the understanding of the phase diagram of strongly interacting 
matter described by Quantum Chromodynamics (QCD), much of it driven from the lattice studies. The 
Equation of state (EoS) relating the energy density and pressure to temperature is known very precisely from lattice 
studies in the continuum limit for vanishing baryon density \cite{Borsanyi:2013bia, Bazavov:2014pvz} and has been extended to 
finite baryon densities as well \cite{Allton:2002zi,Gavai:2003mf,Allton:2003vx,Allton:2005gk,Ejiri:2005uv,Borsanyi:2012cr}. At low 
temperatures the thermodynamic quantities are found to be in quite good agreement with the hadron 
resonance gas (HRG) model approximation, although systematic deviations have been observed, 
which may be due to the existence of additional resonances which are not taken into 
account in HRG model calculations based on experimentally measured resonances 
\cite{Majumder:2010ik,Bazavov:2014xya}. Beyond the fundamental understanding of the non-perturbative aspects of
QCD medium at finite temperature, the EoS at vanishing chemical potentials is an essential input 
into the hydrodynamic modeling of hot and dense matter created in heavy ion 
collisions at the LHC and the highest RHIC beam energies. However BES run I and the upcoming 
BES-II experiments at RHIC will be probing hot and dense media characterized by $0\leq \mu_B/T \leq\ 3$ 
if indeed thermalization is achieved under experimental conditions. Thus lattice input of the EoS at 
non-vanishing baryon number, strangeness and electric charge chemical potentials 
is a crucial theoretical input for the upcoming experimental efforts. Due to the sign problem 
a direct calculation of the EoS at non-zero $(\mu_B,\ \mu_Q,\ \mu_S)$ using lattice techniques 
is unfortunately not yet possible. This problem can be circumvented at small enough densities by performing
Taylor expansion of the thermodynamic quantities \cite{Allton:2002zi,Gavai:2001fr}. Moreover the presence 
of non-analyticities in the phase diagram e.g., a possible critical end-point, can be \cite{CEP,Stephanov} also located 
from the radius of convergence of the Taylor series of the baryon number fluctuation~\cite{Gavai:2004sd}. In order to have a control 
on the Taylor expansion for pressure for a wide range of $\mu_B$ and to study its radius of convergence, it is important 
to measure the higher order expansion coefficients which are simply the generalized baryon number susceptibilities. 

In this proceedings we report our latest results for the Taylor coefficients of pressure and baryon number density upto sixth order in $\mu_B$~\cite{Bazavov:2017dus}. This enables 
us to calculate the EoS in the continuum limit for $\mu_B/T< 2.5$ which is expected to describe heavy ion collisions at centre of mass energies as low as $\sqrt s\sim 11 GeV$~\cite{Cleymans:2005xv}. 
We also provide a new estimate for the location of the critical end-point in the QCD phase diagram.

\section{The QCD Equation of state in QCD under strangeness neutral conditions}
To describe the medium created in heavy ion collisions one has to impose strangeness 
neutrality condition i.e., $n_S=0$. Moreover for most colliding heavy ions, there is a 
constraint on the net baryon to charged species i.e., $n_Q/n_B=0.4$~\cite{Bazavov:2012vg}. These conditions constrain 
the values of $\mu_B, \mu_Q, \mu_S$,  that enter into the calculations of the EoS.
We expand $\mu_S$ and $\mu_Q$ as a series in $\mu_B$ so that the derivatives of the partition 
function with respect to $\mu_S$ and $\mu_Q$ can be expressed in terms of $\mu_B$. The details of 
the derivation of thermodynamic quantities like pressure, energy density for the constrained case 
are given in \cite{Bazavov:2017dus}. 

\begin{figure}[t]
\includegraphics[scale=0.4]{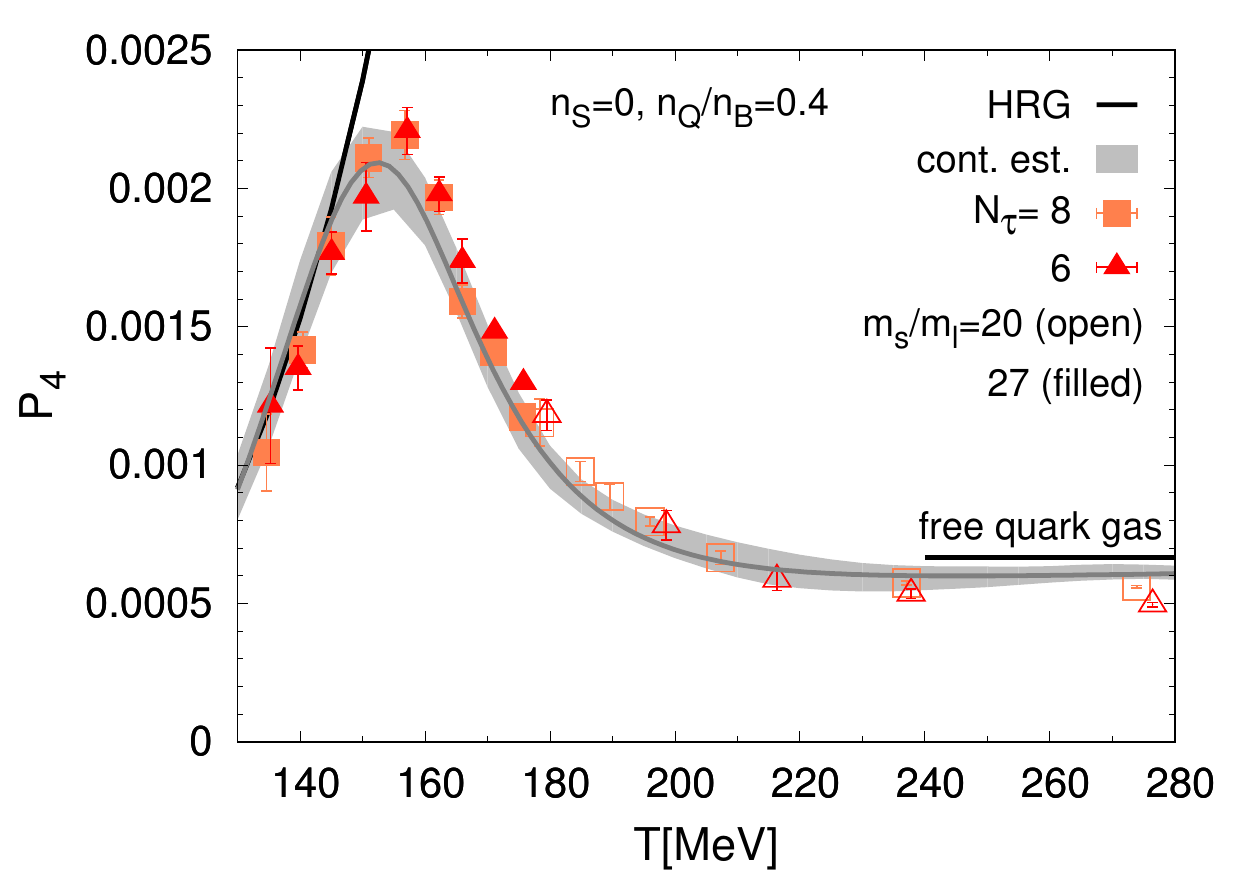}
\includegraphics[scale=0.4]{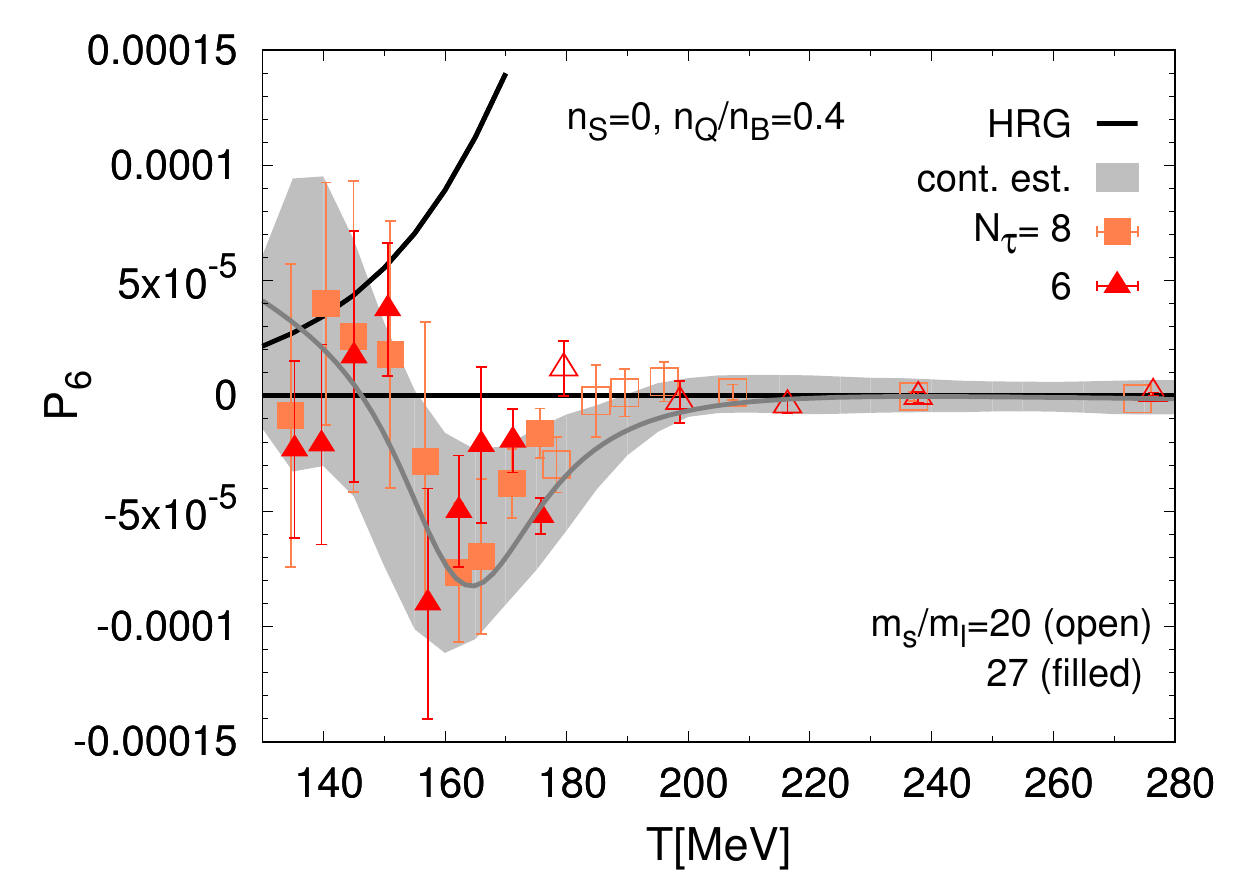}
\hspace{-0.8cm}
\includegraphics[scale=0.5]{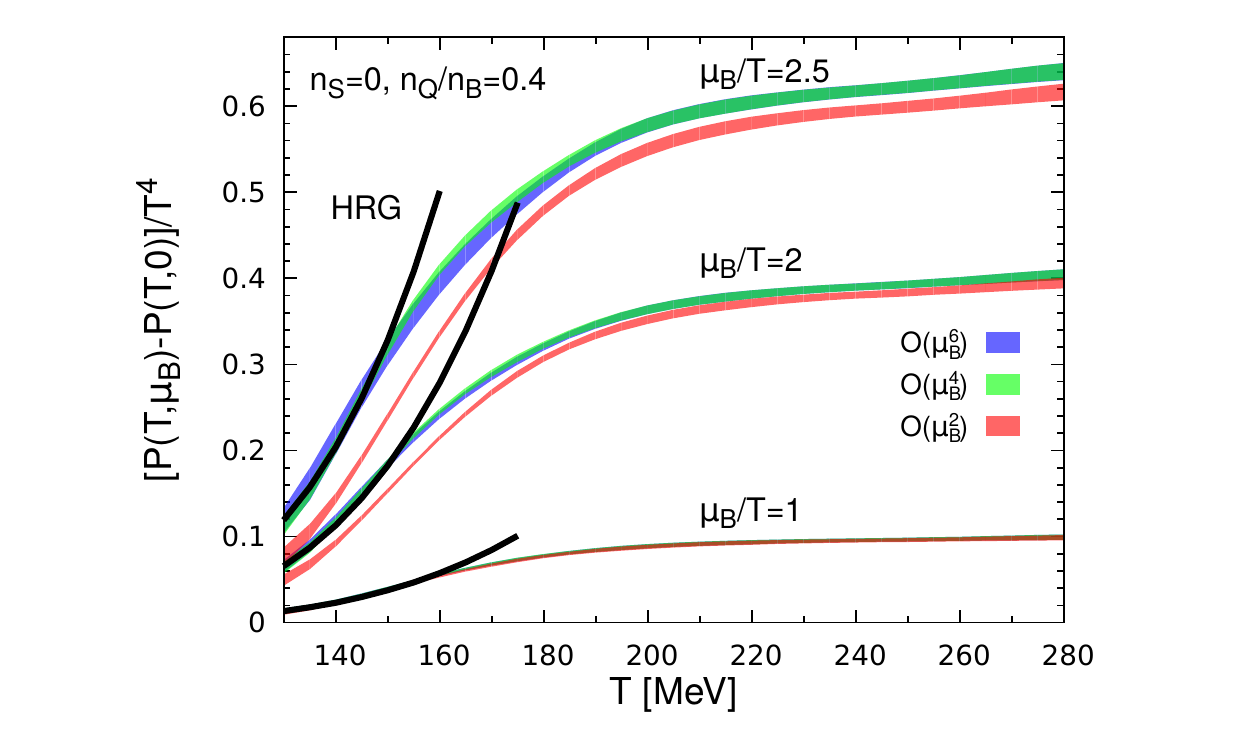}
\caption{The fourth $P_4$ and sixth order, $P_6$ Taylor coefficients for the pressure in 2+1 flavor QCD under the strangeness neutrality 
constraint $n_S=0$ and corresponding to  $n_B/n_Q=0.4$ are shown in the left and the center panels.
The net pressure for different values of $\mu_B/T$ as a function of temperature for the same constraint conditions is shown in the 
rightmost panel. All of the figures are taken from \cite{Bazavov:2017dus}. }
\label{fig:Pseries}
\end{figure}

Our results are based on the lattice QCD calculation with 2+1 flavors using Highly Improved Staggered Quark (HISQ) discretization 
scheme. We have chosen different lattices of extent $N_s^3\times N_\tau$ where $N_\tau=6,8,12,16$ and fixed $N_s=4 N_\tau$ 
in order to be in the thermodynamic limit. The strange quark mass are always tuned to its physical value. The light quark masses are 
chosen such that the Goldstone pion mass $m_\pi=140$ MeV for $T\leq 175$ MeV. At high temperatures since the thermodynamic quantities are not sensitive to the 
choice of the light quark masses we have chosen heavier than physical light quark mass which corresponds to $m_\pi=160$ MeV. 
For further technical details see Ref.~\cite{Bazavov:2017dus}. We expand the pressure in the constrained case as a series in  $\mu_B/T$, as 
\begin{equation}
 \frac{P(T,\mu_B)}{T^4}=\frac{P(T,\mu_B=0)}{T^4}+ P_2 \left(\frac{\mu_B}{T}\right)^2+ P_4 \left(\frac{\mu_B}{T}\right)^4+ 
 P_6 \left(\frac{\mu_B}{T}\right)^6+...~,~P_n=\frac{1}{n! VT^3}\frac{\partial^n ln \mathcal Z_{QCD}}{\partial  \left(\frac{\mu_B}{T}\right)^n}~.
\end{equation}
The computational complexity of the lattice calculation of the coefficients $P_n$ increases with each order $n$. For $n=2,4$ we use the 
conventional method of introducing $\mu_B$ which is known to cancel any unphysical divergences explicitly for $n\leq 4$ 
~\cite{Hasenfratz:1983ba}. 
However for $n\geq 4$ we adopt a new method of introducing $\mu$ which significantly reduces the 
computational costs~\cite{Gavai:2011uk} for these higher order $P_n$ without introducing any potentially divergent term~\cite{Gavai:2014lia}. 
Our results for Taylor coefficients for pressure are shown in the left and central panels of Fig. \ref{fig:Pseries} for different lattice spacings. 
The gray band represents the continuum extrapolated results obtained from our $N_\tau=6,8$ data.  For $P_4$ we observe a deviation from the HRG values already 
for $T\geq 150$ MeV. For $P_6$ the central values are systematically below the HRG estimates for $T>145$ MeV however given the current 
errors we cannot make a conclusive statement. However with the high statistics data, we can clearly observe a negative dip in $P_6$ immediately 
in the vicinity  and above the chiral crossover region given by $T_c=154(9)$ MeV~\cite{Bazavov:2011nk}. This is a genuine signature of the non-perturbative 
nature of the QCD medium just above deconfinement which cannot be reproduced within Hard Thermal Loop perturbation theory~\cite{Haque:2014rua}. Our results 
are in good agreement with the continuum estimates obtained from imaginary $\mu$ techniques using analytic continuation \cite{Gunther:2016vcp,BorsanyiQM}.

Using these estimates of $P_n$, we have calculated the pressure in QCD for different values of $\mu_B/T$ relevant for RHIC BES 
experiments, the continuum estimates of which are shown in the rightmost panel of Fig. \ref{fig:Pseries}. For $T>170$ MeV we infer that it is sufficient to consider
terms upto $\mathcal O(\mu_B^6)$ for calculating pressure for a wide range of $\mu_B/T \leq 2.5$ due to convergence of the results at this order 
compared to the results obtained from truncation at $\mathcal O(\mu_B^4)$. For $T<165$ MeV, the convergence is still fairly good limited only due to the 
large errors in the $P_6$ data. For the net baryon number density shown in the left panel of Fig. \ref{fig:Nseries}, terms upto 
$\mathcal O(\mu_B^6)$ are not sufficient for its estimation for $\mu_B/T> 2$ as we do not find a convergence of the results from 
different orders of truncation. This is  due to the fact that the higher order Taylor coefficients contribute with a larger numerical prefactor 
in the baryon number density compared to pressure. Hence a precise calculation of the sixth order coefficient and extending to further higher orders are needed 
for estimating the baryon number density for $\mu_B/T>2$. In summary we now have continuum extrapolated results for the EoS for $\mu_B/T\leq 2$ for a wide range 
of temperatures. A parameterization of the pressure and number densities as a function of $T_c/T$ for the constrained case are available, for more details 
see \cite{Bazavov:2017dus}.

\section{The critical end-point estimates upto $\mathcal O(\mu_B^6)$}
\begin{figure}[t]
\includegraphics[scale=0.55]{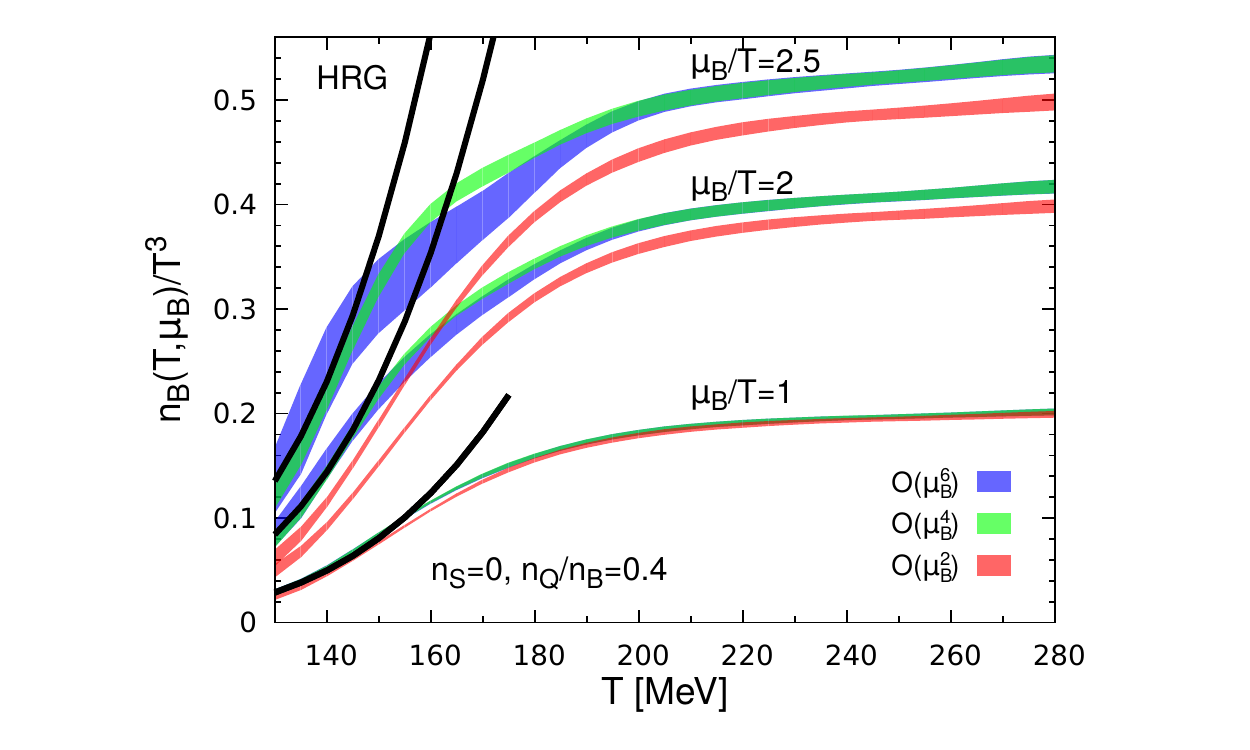}
\includegraphics[scale=0.45]{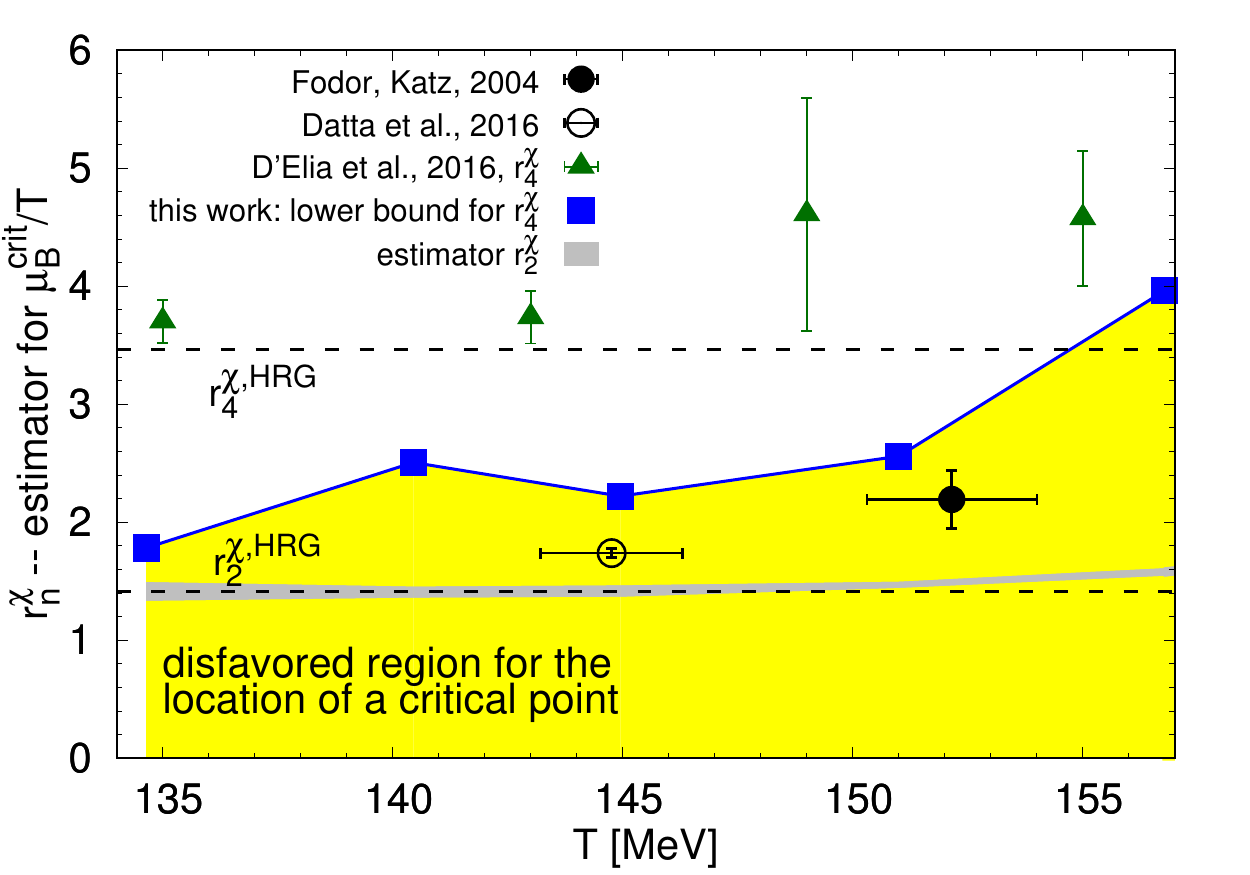}
\caption{The net baryon number density for different $\mu_B/T$ as a function of temperature is shown in the left panel from ~\cite{Bazavov:2017dus}. 
Our results for the estimators for the radius of convergence of Taylor
series for $\chi_2^B(T,\mu_B)$, for $\mu_S=\mu_Q=0$ and $N_\tau =8$ are shown in the 
right panel. Our results are compared 
with the calculations with an imaginary chemical potential (triangles) \cite{DElia:2016jqh},
using a reweighting technique \cite{Fodor:2004nz} and Taylor expansions \cite{Datta:2016ukp} for 
standard staggered fermion discretization rescaled using $T_c=154$~MeV taken from ~\cite{Bazavov:2017dus}.
}
\label{fig:Nseries} 
\end{figure}

The net baryon-number susceptibility, $\chi_2^B$ is expected to diverge at the critical end-point, so the ratios of the successive terms in its 
Taylor series, would give an estimate for the radius of convergence.  Having obtained the coefficients of the Taylor 
series for pressure, we can calculate estimates for the radius of convergence as 
$r_{2n}^{\chi} = \left| \frac{2n (2n-1)P_{2n}}{P_{2n+2}(2n+2)(2n+1)} \right|^{1/2}$, but for $\mu_S=\mu_Q=0$. These estimates will 
converge to the true radius of convergence in the limit $n\rightarrow\infty$. Our results for
$r_{2n}^{\chi}$ are shown in the right panel of Fig. \ref{fig:Nseries}. The upper boundary of the excluded 
region is obtained by considering the upper values of errors in our $P_6/P_4$ data.  
Our results are consistent with  the very recent results for these ratios
obtained from calculations with an imaginary $\mu_B$ \cite{DElia:2016jqh}, where 
the estimates of $r_4^{\chi}$ are larger than our current lower bound. Estimators 
calculated based on a reweighting technique \cite{Fodor:2004nz} as well as 
from Taylor series expansion in 2-flavor QCD \cite{Datta:2014zqa,Datta:2016ukp} are consistently 
lower than our bounds. However one has to note that all the lattice results are at fixed lattice spacing 
which leads to the present differences, which hopefully would agree in the continuum limit.

\section{Summary and Outlook}
We now have continuum extrapolated results for the QCD EoS for $\mu_B/T\leq 2$ under the constraints that 
are realized in heavy-ion experiments using Taylor expansion upto $\mathcal O(\mu_B^6)$ . Furthermore 
we have measured different estimators for the radius of convergence of the net baryon number fluctuation, which 
allows us to rule out the existence of a critical end-point in the QCD phase diagram for $\mu_B/T\leq 2$ and
$145\leq T\leq 155$ MeV. This is consistent with the fact that we find no non-analyticities in the Taylor expansion of 
pressure for these values of $\mu_B/T$. In order to extend our results to 
$\mu_B/T>2$ to be accessible in the BES II experiments at RHIC and provide more robust bounds on the location of 
the critical end-point we need to calculate the eighth order Taylor coefficient for pressure, which is well within our reach.

\textbf{Acknowledgements:}
This work was supported in part through Contract No. DE-SC001270 with the U.S. Department of Energy, through
the Scientific Discovery through Advanced Computing (Scidac) program funded by the U.S. Department of Energy,
Office of Science, Advanced Scientific Computing Research and Nuclear Physics, the DOE funded BEST topical
collaboration, the NERSC Exascale Application Program (NESAP), the grant 05P12PBCTA of the German Bundesministerium f\"{u}r 
Bildung und Forschung, the grant 56268409 of the German Academic Exchange Service (DAAD),
grant 283286 of the European Union and the National Natural Science Foundation of China under grant numbers
11535012 and 11521064. Numerical calculations have been made possible through an INCITE grant of USQCD, ALCC
grants in 2015 and 2016, and PRACE grants at CINECA, Italy, and the John von Neumann-Institute for Computing
(NIC) in Germany. These grants provided access to resources on Titan at ORNL, Bluegene/Q at ALCF and NIC,
Cori I and II at NERSC and Marconi at CINECA. Additional numerical calculations have been performed on USQCD
GPU and KNL clusters at Blab and Fermilab, as well as GPU clusters at Bielefeld University, Paderborn University,
and Indiana University.




\bibliographystyle{elsarticle-num}
\bibliography{<your-bib-database>}







\end{document}